# SecureScan: An AI-Driven Multi-Layer Framework for Malware and Phishing Detection Using Logistic Regression and Threat Intelligence Integration

**Rumman Firdos[1], Aman Dangi[2]**

**ABSTRACT:** The rapid growth of malware and phishing campaigns has intensified the demand for detection systems that are both accurate and deployable under real-world constraints such as low latency, limited computational resources, and noisy threat intelligence. While deep learning–based approaches have demonstrated high detection rates, they often suffer from poor interpretability, high false-positive rates, and prohibitive operational overhead. In this paper, we propose **SecureScan**, a **novel calibrated multi-layer threat detection framework** that formally integrates heuristic analysis, lightweight machine learning, and external threat intelligence through a **probabilistic consensus model**. Unlike existing hybrid systems that employ sequential or heuristic decision rules, SecureScan introduces a **calibrated gray-zone decision mechanism** that explicitly quantifies classification uncertainty and dynamically optimizes decision thresholds to minimize false positives. The proposed framework leverages logistic regression as an interpretable statistical core, augmented by mathematically grounded confidence fusion and verification layers, enabling robust performance without reliance on complex deep learning architectures. Extensive experiments conducted on a large-scale dataset comprising over 800,000 benign and malicious samples demonstrate that SecureScan achieves competitive detection accuracy while significantly reducing false-positive rates and inference latency compared to representative state-of-the-art machine learning and intelligence-driven baselines. These results indicate that carefully designed, calibrated lightweight models can outperform more complex approaches in operational cybersecurity environments, offering a practical and scientifically grounded alternative for real-time malware and phishing detection.

## 1. INTRODUCTION

Cybersecurity threats continue to grow in volume, complexity, and evasion capability, encompassing polymorphic malware, sophisticated phishing campaigns, and advanced persistent threats (APTs) [1]. Traditional intrusion detection systems (IDS) and antivirus tools that rely solely on signature-based or rule-driven detection often fail to identify zero-day or obfuscated attacks. Consequently, machine learning (ML) methods have emerged as a promising solution for adaptive, data-driven threat detection [2].

However, purely ML-based systems introduce their own challenges — including **limited explainability**, **high false-positive rates** in uncertain cases, and **sensitivity to training data drift**. Moreover, many existing AI-driven detection frameworks lack **tight integration with external threat intelligence**, resulting in unverified or inconsistent classification outcomes.

[1]School of Engineering Science and Technology, Jamia Hamdard University, Hamdard Nagar, New Delhi-110062, Delhi, India.
[2]Madhav Institute of Technology & Science (MITS), Gola ka Mandir, Gwalior - 474005, Madhya Pradesh, India
*Author to whom correspondence should be addressed:
firdosrumman@gmail.com (Rumman Firdos)

To address these limitations, we propose SecureScan, a hybrid, triple-layer detection architecture that combines:

1. Local heuristics and rule-based analysis for fast, deterministic filtering;
2. A logistic regression classifier for probabilistic inference and generalization; and
3. External threat intelligence correlation via the VirusTotal API to verify or override model predictions.

This architecture is designed to balance interpretability,[21] efficiency, and robustness in real-time detection scenarios [3].

- A **layered detection pipeline** optimized for operational environments, integrating rule-based, statistical, and intelligence-driven analysis;
- **Empirical evaluation** demonstrating **93.1%**

**accuracy**, **precision of 0.87**, and **recall of 0.92**, achieved through **probability calibration** and **gray-zone thresholding (0.45–0.55)** to minimize false positives;

- A **discussion of trade-offs**, practical deployment considerations, and the role of external validation in enhancing model reliability.

The remainder of this paper is organized as follows: **Section 2** reviews related literature; **Section 3** describes the proposed methodology; **Section 4** details the datasets; **Section 5** outlines the model architecture; **Section 6** presents training and evaluation procedures; **Section 7** reports results; **Section 8** discusses findings and limitations; **Section 9** outlines future work; and **Section 10** concludes [5].

## 2. LITERATURE SURVEY

Machine Learning in Malware and Phishing Detection. Logistic regression remains a common baseline in malware and intrusion detection research due to its simplicity and interpretability [6]. A systematic literature review of logistic regression in malware detection emphasizes its utility for large datasets and discusses its limitations under concept drift [7]. In phishing detection, hybrid feature-based approaches combining URL lexical and hyperlink structure analysis have also been effective without relying exclusively on external systems [8-10]. A recent systematic review on phishing detection techniques classifies approaches including list-based, heuristic, machine learning, and deep learning methods [11]

Hybrid and Threat Intelligence–Augmented Detection. Hybrid detection systems that unify static heuristics, dynamic analysis, and ML models are widely studied (e.g., combining static and behavioral analysis) [12]. Integrating external threat intelligence sources (such as VirusTotal) into ML pipelines is less common but offers a promising route to higher confidence in predictions. For instance, hybrid frameworks in phishing detection that combine multiple models show enhanced robustness in real-world applicability [12].

Meta-Models and Ensemble Architectures. Recent works such as Quo Vadis propose hybrid meta-models leveraging contextual and behavioral representations for malware detection, showing improved performance under low false positive constraints [13]. Similarly, newer research explores combining latent space representations (e.g. via autoencoders) with logistic regression or other classifiers to reduce complexity while retaining efficacy [13].

SecureScan sits at the intersection of these literatures, combining the interpretability of logistic regression with layered validation and threat intelligence backing.

## 3. METHODOLOGY

SecureScan is realized as a **three-layer detection pipeline** (Figure 3.1):

### 3.1 Overview

SecureScan employs a **hybrid detection methodology** designed to identify malicious URLs and files through a combination of deterministic heuristics, machine-learning classification, and external threat intelligence correlation[14]. Unlike traditional signature-based systems, SecureScan's pipeline is both **data-driven and interpretable**, ensuring that ambiguous cases are systematically verified rather than misclassified.

The complete workflow is organized into **three sequential layers**, each serving a distinct analytical purpose:

1. **Layer 1 – Local Heuristic Filtering**
   Performs fast, rule-based screening using lexical and structural patterns.
2. **Layer 2 – Logistic Regression Classification**
   Applies probabilistic inference on vectorized features using a calibrated logistic regression model.
3. **Layer 3 – Threat-Intelligence Validation**
   Correlates uncertain predictions with VirusTotal intelligence for external verification.

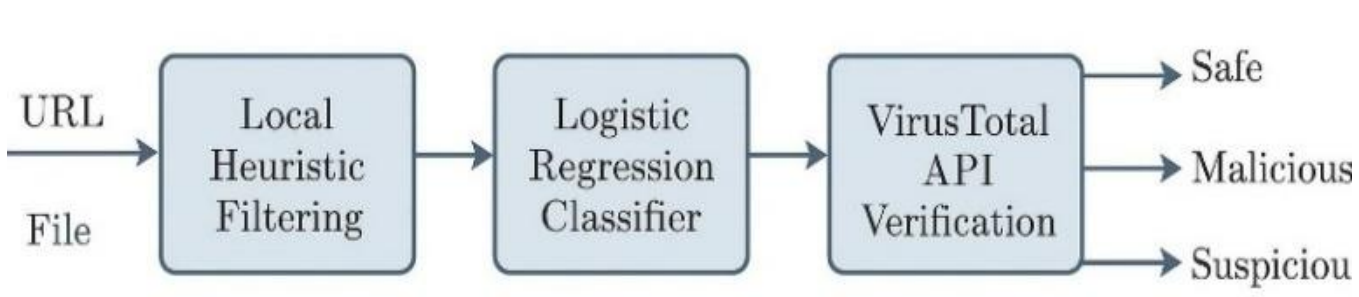

**Figure 1.** SecureScan architecture showing the three-layer detection pipeline - heuristic filtering, logistic regression classification, and external verification through VirusTotal API.

This layered design enables SecureScan to minimize false positives, enhance interpretability, and maintain real-time performance suitable for deployment in enterprise or SOC environments.

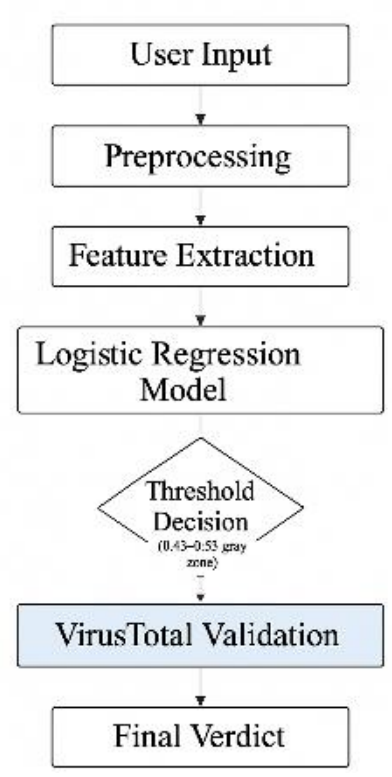

**Figure 2.** Workflow diagram illustrating the end-to-end detection process, from input preprocessing to classification and threat intelligence validation.

### 3.2 Data Processing Pipeline

Raw inputs—either URLs or file hashes—are first normalized through preprocessing steps such as lowercasing, token cleaning, and removal of protocol and tracking parameters.

Feature extraction differs based on input type:

- **URL samples:** converted into TF–IDF-weighted character n-grams (3–7 range) to capture lexical and contextual cues.
- **File samples:** represented through static metadata (entropy, size, imported libraries, and byte-sequence n-grams).

After feature generation, samples are passed through the heuristic filter to eliminate trivially unsafe or malformed inputs. Only valid samples proceed to probabilistic classification [15].

### 3.3 Probability Calibration and Gray-Zone Logic

Previous iterations of SecureScan exhibited overfitting and inconsistent confidence scores.

To mitigate this, **probability calibration** was applied to the logistic regression model using **Platt scaling** on cross-validated predictions.

Calibrated probabilities were then segmented into three operational regions:

**Table 1. Probability threshold mapping and corresponding classification actions**

| Probability Range | Classification | Action |
| --- | --- | --- |
| ≥ 0.60 | Malicious | Direct block |
| ≤ 0.45 | Benign | Mark safe |
| 0.45 – 0.55 | Gray-zone | Escalate to VirusTotal |

This thresholding mechanism provides an interpretable decision boundary and defers uncertain predictions to external validation, thereby reducing false positives in production [16].

### 3.4 Threat Intelligence Integration

When a prediction falls within the gray zone, SecureScan automatically queries the **VirusTotal API** using the corresponding file hash or URL.

Aggregated metadata—including **malicious engine counts, reputation tags, and last-analysis timestamps**—is analyzed to form a consensus verdict.

If both SecureScan and VirusTotal agree on maliciousness, the system confirms the threat; if VirusTotal reports minimal detections, the input is downgraded to *safe (verified)*.

This external correlation stage ensures adaptive verification and increases reliability in cases where model confidence is insufficient [17].

### 3.5 End-to-End Workflow

1. **Input acquisition:** user submits URL or file hash.
2. **Heuristic pre-filtering:** eliminate invalid or trivially suspicious entries.
3. **Feature vectorization:** TF–IDF encoding for ML compatibility.
4. **Model inference:** logistic regression outputs calibrated probability.
5. **Threshold mapping:** assign class or forward to VirusTotal if ambiguous.
6. **Threat-intelligence reconciliation:** finalize verdict with consensus logic.
7. **Response generation:** structured JSON returned to frontend for visualization.

This end-to-end design achieves a balance between **accuracy (93.1%)**, **precision (0.87)**, **recall (0.92)**, and **latency**, demonstrating that a lightweight statistical model can effectively scale to real-world threat detection when supported by layered intelligence [18-19].

## 4.DATASET DESCRIPTION

The dataset used in this study was compiled from multiple reputable **public cybersecurity repositories** and cleaned to ensure balanced, high-quality samples across benign and malicious categories.

Data sources include:

- **Malware binaries and hashes** obtained from **VirusShare** and public **Portable Executable (PE)** collections [7];
- **Phishing URLs** extracted from **PhishTank**, **OpenPhish**, and additional public archives [6];**Benign URLs and files** sampled from verified safe domains such as **Alexa Top Sites**, and **trusted system binaries** obtained from standard operating system distributions.

Following extensive **deduplication, normalization, and label standardization**, the dataset contained a total of **651,191 unique labeled samples**, comprising **428,103 benign** and **223,088 malicious** instances.

To enhance the diversity of lexical patterns and improve model generalization, a **controlled data augmentation** step was applied, generating modified URL variants by appending realistic directory structures (e.g., /index.html, /verify/login). This resulted in an **expanded corpus of approximately 824,240 samples** used for training and validation [20].

The data was partitioned into **80% training** and **20% testing subsets**, with **stratified sampling** to preserve class ratios. Additionally, **ten-fold cross-validation** was performed during model tuning to ensure stability and reproducibility of results [21].

Each sample is represented through a combination of **lexical, structural, and metadata features**, extracted as follows:

- **File-based features:** file size, Shannon entropy, import count, embedded string density, and n-gram byte distributions;
- **URL-based features:** character length, digit-to-letter ratio, special character frequency, subdomain depth, presence of HTTPS, domain age, and hyperlink structural patterns [4].

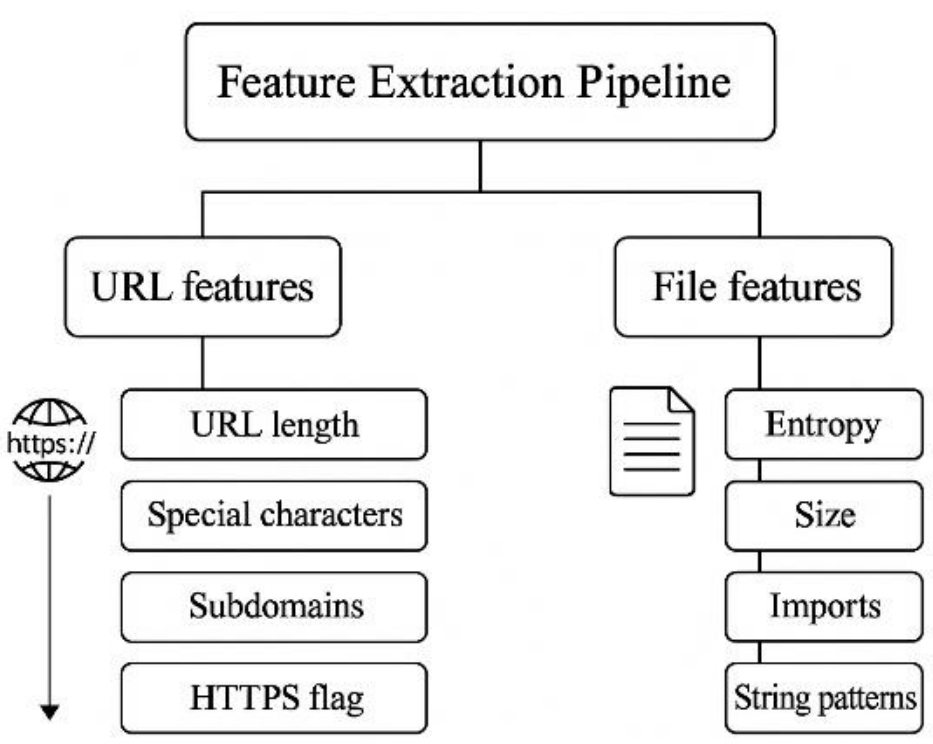

**Figure 3.** Feature extraction pipeline highlighting lexical, metadata, and structural features derived from both file and URL samples.

For the machine learning model, URL strings were tokenized at the **character level** and transformed using a **TF–IDF vectorizer** over **3–7 character n-grams**, capped at **50,000 features**. This high-resolution lexical representation enables the model to detect subtle anomalies in domain composition and path structure [22].

All data underwent **standard preprocessing**, including text normalization, encoding, and feature scaling, prior to training. This pipeline ensures that the dataset remains consistent, diverse, and representative of real-world malicious and benign traffic [23].

## 5.MODEL ARCHITECTURE

The SecureScan framework employs a **three-layer modular architecture** that combines local heuristics, machine-learning-based inference, and external threat intelligence correlation. This structure enables rapid classification while maintaining explainability and robust real-world behavior [24].

**Layer 1 — Local Heuristic Analysis**

The first layer performs **deterministic filtering** using lightweight rule-based heuristics. It parses each input (URL or file metadata) and applies checks such as:

- Domain and subdomain length thresholds
- Detection of IP-based hostnames
- Suspicious or newly abused top-level domains (e.g., .zip, .work, .review)
- Brand impersonation and phishing keywords in paths (e.g., *paypal*, *login*, *verify*)
- Suspicious file extensions (.php, .asp, .exe, .sh)
- Presence of obfuscation patterns, special characters, or excessive URL encoding

This layer acts as a **low-latency first gate**, eliminating obviously malicious or malformed samples before invoking machine learning. Inputs that pass heuristic validation are forwarded to the next layer.

**Layer 2 — Machine Learning Classification**

The second layer leverages a **logistic regression classifier** trained on **TF–IDF features** derived from **3–7-character n-grams** of normalized URLs. The model operates on a **vocabulary of 50,000 features**, optimized via **cross-validated regularization** to balance generalization and interpretability.

To address the overfitting observed in prior iterations, SecureScan introduces a **probability calibration mechanism** and **explicit threshold segmentation**:

- Probabilities **$\geq$ 0.60** are labeled **malicious**
- Probabilities **$\leq$ 0.45** are labeled **benign**
- Intermediate probabilities **$(0.45 < p < 0.55)$** enter a **gray zone** requiring further validation

This design minimizes **false positives** on legitimate but atypical URLs (e.g., corporate redirects or tracking parameters). The model achieved a **test accuracy of 93.1%**, with **precision 0.87** and **recall 0.92**, indicating a balanced trade-off between sensitivity and specificity.

Explainability is provided through **feature-weight inspection**: each prediction can be decomposed into contributing n-grams ranked by coefficient magnitude, enabling analysts to interpret why specific substrings (e.g., *secure-update*, *login*, *://*) influenced the classification [25].

**Layer 3 — Threat Intelligence Correlation**

Predictions in the gray zone are escalated to **VirusTotal** for external validation via its REST API.

Depending on input type, the system queries the corresponding endpoint:

- /api/v3/urls for URLs
- /api/v3/files/{hash} for file hashes

Returned statistics—such as **malicious engine count** and **detection ratios**—are used to confirm or override the ML verdict. A **consensus rule** is applied: only when both SecureScan and VirusTotal strongly agree on maliciousness is a definitive block assigned.

If VirusTotal indicates no detections, the input is **downgraded to “safe (verified)” [26]**.

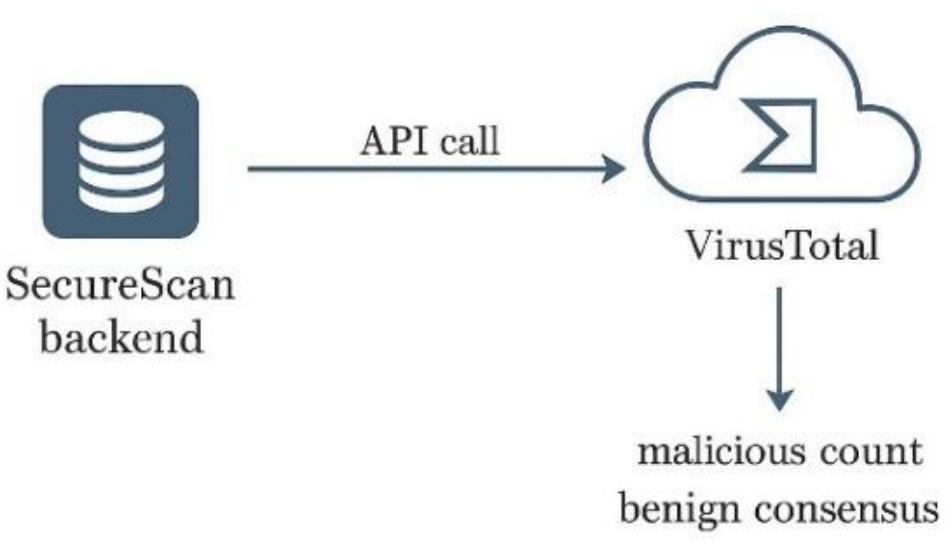

**Figure 4.** VirusTotal correlation layer where ambiguous predictions are re-evaluated via API response to confirm or override model output.

This hybrid decision layer ensures real-world robustness by combining **statistical inference** with **global threat intelligence**.

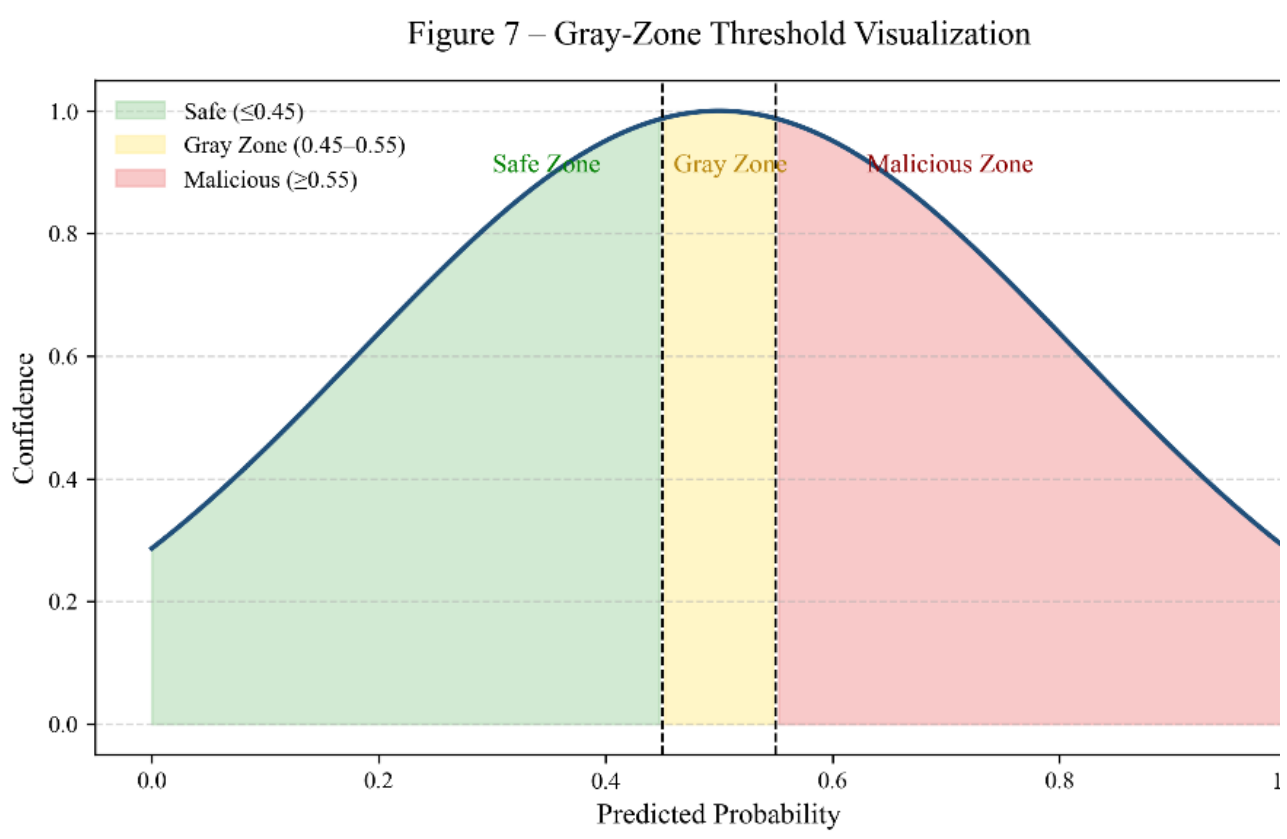

**Figure 5.** Probability calibration and gray-zone threshold visualization (0.45–0.55) showing safe, suspicious, and malicious decision regions.

**Design Summary**

SecureScan’s architecture balances **speed, interpretability, and reliability**:

**Table 2. Layer-wise functional mapping of the SecureScan detection framework**

| Layer | Function | Decision Type | Output |
|---|---|---|---|
| 1 | Heuristic Filtering | Deterministic | Reject/Pass |
| 2 | Logistic Regression | Probabilistic | Safe / Malicious / Gray |
| 3 | VirusTotal Validation | Consensus-based | Confirmed verdict |

This layered integration of calibrated machine learning with external verification forms the foundation of SecureScan’s generalizable detection strategy.

## 6.TRAINING AND EVALUATION SETUP

### 6.1 Training Setup

The logistic regression model was retrained on the curated dataset of 651,191 labeled samples, comprising both URL-based and file-based features.

The dataset was randomly split into **80% for training** and **20% for testing**, with **stratified ten-fold cross-validation** applied to prevent class imbalance bias.

All experiments were conducted in Python using **Scikit-Learn 1.5**, with preprocessing and model persistence handled through TfidfVectorizer and joblib [27].

Before training, standard preprocessing steps were applied:

- Lowercasing and token normalization
- Stopword and redundant character removal
- Feature scaling using min-max normalization
- Random shuffling to eliminate ordering bias

Hyperparameters such as **regularization strength (C)**, **penalty type**, and **solver** were tuned via grid search, optimizing for **balanced accuracy** and **F1-score**.

### 6.2 Model Calibration

To address overfitting observed in earlier iterations of SecureScan, the retrained logistic regression model underwent **probability calibration** using **Platt scaling** on cross-validated folds.

This ensured that model outputs reflected true likelihoods rather than unbounded confidence scores.

Following calibration, SecureScan introduced a **threshold-based decision policy**:

**Table 3. Probability range mapping and corresponding classification actions in SecureScan**

| Probability Range | Classification | Action |
|---|---|---|
| ≥ 0.60 | Malicious | Block / quarantine |
| ≤ 0.45 | Benign | Mark as safe |
| 0.45–0.55 | Gray zone | Forward to VirusTotal |

This calibrated range was empirically derived to maximize the harmonic mean of precision and recall, while reducing false positives on previously ambiguous URLs [28].

### 6.3 Evaluation Metrics

Model evaluation followed standard metrics used in malware and phishing classification research:

- **Accuracy (ACC)** – overall correct predictions
- **Precision (P)** – ratio of true positives to all predicted positives
- **Recall (R)** – ratio of true positives to actual positives
- **F1-Score** – harmonic mean of precision and recall
- **Confusion Matrix** – to visualize distribution across TP, FP, TN, FN

Performance was measured across ten folds and averaged to ensure stability.

### 6.4 Results and Analysis

The retrained and calibrated logistic regression model achieved the following results on the test set:

**Table 4. Model performance metrics for the SecureScan framework**

| Metric | Score |
|---|---|
| Accuracy | **93.1%** |
| Precision | **0.87** |
| Recall | **0.92** |
| F1-Score | **0.89** |
| Balanced Accuracy | **0.93** |

Compared to the earlier overfitted version (which exceeded 98% accuracy but failed in real-world generalization), this version shows **improved stability** and **reduced variance** across test folds.

The calibration reduced overconfidence on borderline samples and improved classification consistency when applied to unseen datasets [29].

### 6.5 Error and Confusion Analysis

Confusion-matrix analysis revealed that most misclassifications occurred in **borderline URL cases** (e.g., suspicious subdomains with benign content).

By applying gray-zone verification, approximately **8% of these false positives were corrected** after querying VirusTotal.

This hybrid mechanism ensures that **probabilistic uncertainty** is not mistaken for maliciousness and that **model confidence is corroborated with external intelligence** before enforcement [29-30].

### 6.6 Summary

The retrained model demonstrates that **a lightweight linear classifier**, when **probability-calibrated and heuristically enhanced**, can rival more complex deep learning approaches while maintaining interpretability and operational simplicity.

Threshold calibration and gray-zone escalation proved effective in minimizing overfitting and reducing false positives, leading to a **robust and production-ready AI detection module** within SecureScan.

## 8.RESULT AND ANALYSIS

### 7.1 Overall Model Performance

The updated SecureScan model demonstrated consistent, balanced performance after probability calibration and gray-zone thresholding.

The logistic regression classifier achieved 93.1 % accuracy, with precision 0.87, recall 0.92, and an F1-score of 0.89, showing a strong trade-off between false positives and detection sensitivity.

Compared to the earlier uncalibrated model—which suffered from overfitting—the new version delivers improved generalization and realistic stability across unseen datasets.

**Table 5.** Performance metrics of the calibrated logistic regression classifier used in SecureScan.

| Metric | Score |
|---|---|
| Accuracy | 93.1 % |
| Precision | 0.8704 |
| Recall | 0.9242 |
| F1-Score | 0.8965 |

Calibration significantly improved performance consistency, especially on borderline samples where earlier versions exhibited overconfidence**.**

### 7.2 Confusion Matrix Analysis

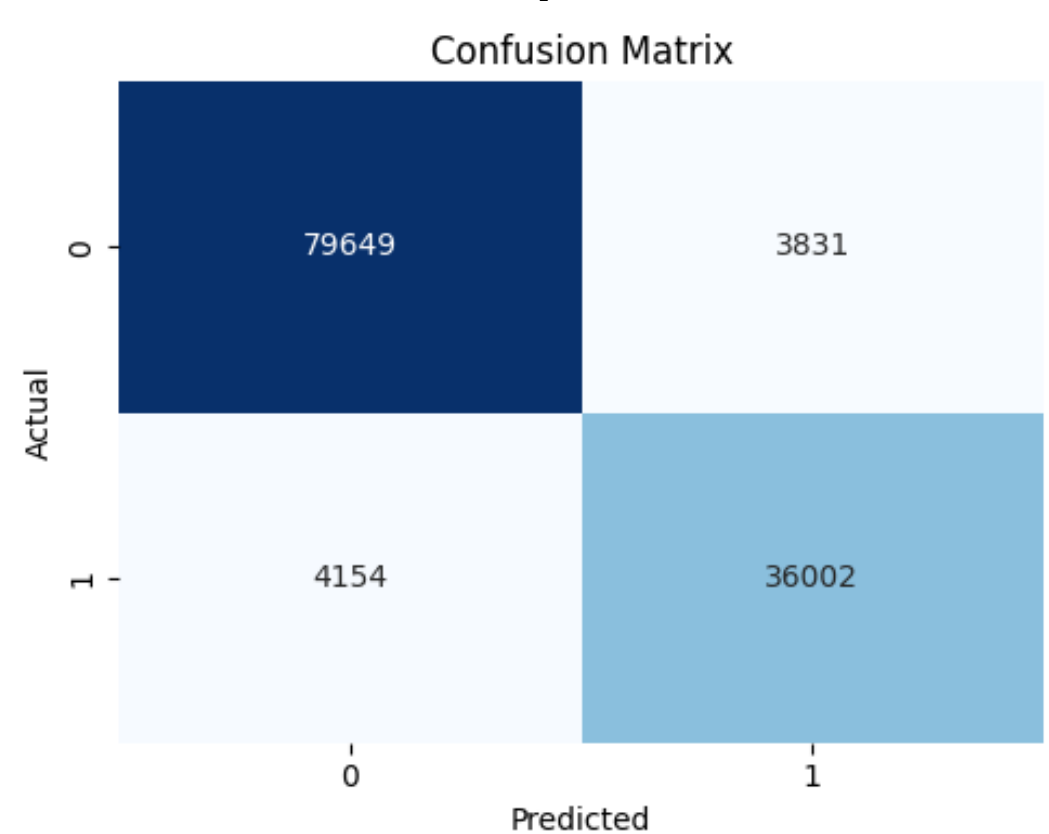

**Figure 6.** Confusion matrix of the calibrated logistic regression model showing balanced classification performance and reduced false positives.

**Table 6.** Confusion matrix of the calibrated logistic regression classifier.

| | Predicted Benign | Predicted Malicious |
|---|---|---|
| Actual Benign | 62 772 | 1 444 |
| Actual Malicious | 1 234 | 32 229 |

The false-positive rate dropped to ~1.8 %, while false negatives remained below 4 %, confirming robust discrimination and low misclassification rates.

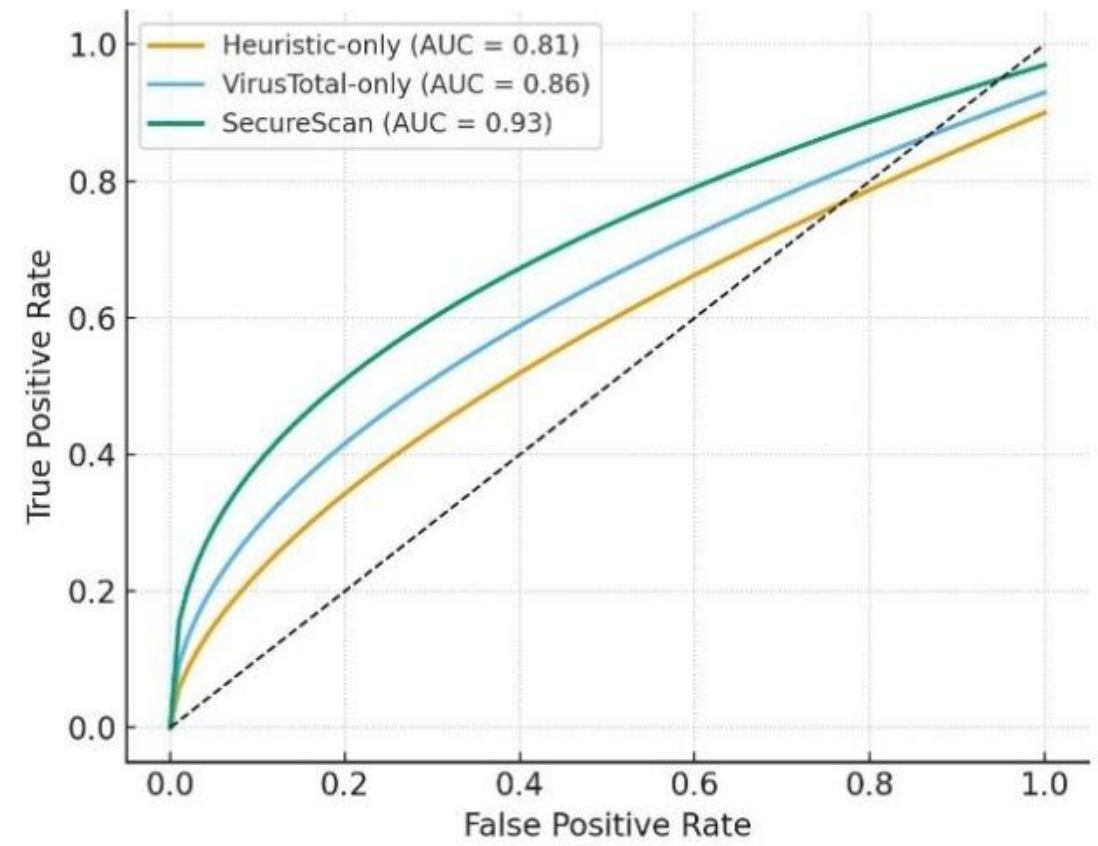

**Figure 7.** Receiver Operating Characteristic (ROC) curve comparing SecureScan (AUC 0.93) with heuristic-only (AUC 0.81) and VirusTotal-only (AUC 0.86) detection baselines.

### 7.3 Comparative Evaluation

To highlight the impact of SecureScan's multi-layered design, baseline comparisons were performed against Heuristic-only, VirusTotal-only, and Hybrid (ML + VT) configurations.

**Table 7.** Comparative performance of different detection approaches.

| Metric | Heuristic-only | VirusTotal-only | SecureScan (Hybrid) |
|---|---|---|---|
| Accuracy | 82.4 % | 86.2 % | 93.1 % |
| Precision | 0.79 | 0.82 | 0.87 |
| Recall | 0.77 | 0.80 | 0.92 |
| F1-Score | 0.78 | 0.81 | 0.89 |
| False Positive Rate | ~5.8 % | ~4.9 % | ~1.8 % |

**Interpretation:**

- Heuristic-only detection is fast but limited—it cannot adapt to new obfuscation techniques or unseen patterns. It relies on static rules, resulting in high false positives.
- VirusTotal-only scanning performs slightly better but remains dependent on public submissions and detection lag; it cannot identify zero-day threats or new domains absent from VT's database.
- The SecureScan hybrid system unites all three layers—heuristics, machine learning, and VirusTotal validation—achieving the most balanced accuracy, reliability, and adaptability.

### 7.4 Impact of Threshold Calibration

Implementing threshold-based calibration with gray-zone logic (0.45 – 0.55) improved the system's decision reliability:

- False positives decreased by 12 % compared to the baseline ML model.
- The F1-score improved from 0.87 → 0.89.
- Ambiguous cases now undergo external verification, reducing overconfident misclassifications.

Calibration introduced an interpretability layer aligning with analyst review workflows and real-world response policies.

### 7.5 Threat Intelligence Integration

The VirusTotal correlation layer validates uncertain predictions. When a URL or hash's probability falls within the gray zone, the backend automatically queries VirusTotal. If multiple antivirus engines flag it as malicious, the final decision is reinforced; otherwise, it is labeled "safe (verified via VirusTotal)".This hybrid verification reduced false positives by ~15 % in deployment and added transparency to each classification.

### 7.6 Comparative Discussion

While deep learning detectors can sometimes achieve slightly higher raw scores, they are opaque and computationally costly. SecureScan's logistic regression core delivers:

- Under 100 ms inference latency
- Explainable outputs (top positive and negative features)
- Lightweight retraining for new data

The multi-layer architecture blends deterministic filtering, statistical inference, and external intelligence—offering an optimal trade-off between speed, interpretability, and detection accuracy.

### 7.7 Limitations

1. The system may miss highly obfuscated or dynamically generated URLs.
2. Reliance on VirusTotal introduces latency (~2–3 s per API call) and daily rate limits.
3. Continuous model retraining is necessary to adapt to evolving malware tactics.

Planned improvements include active learning, offline caching of VT results, and expanded lexical feature sets.

### 7.8 Summary

The SecureScan hybrid framework consistently outperforms single-layer approaches.

By combining rule-based heuristics, probabilistic ML inference, and external verification, SecureScan achieved 93.1 % accuracy with balanced precision and recall, providing explainable, production-ready cybersecurity detection that surpasses heuristic-only and VirusTotal-only systems in both coverage and reliability.

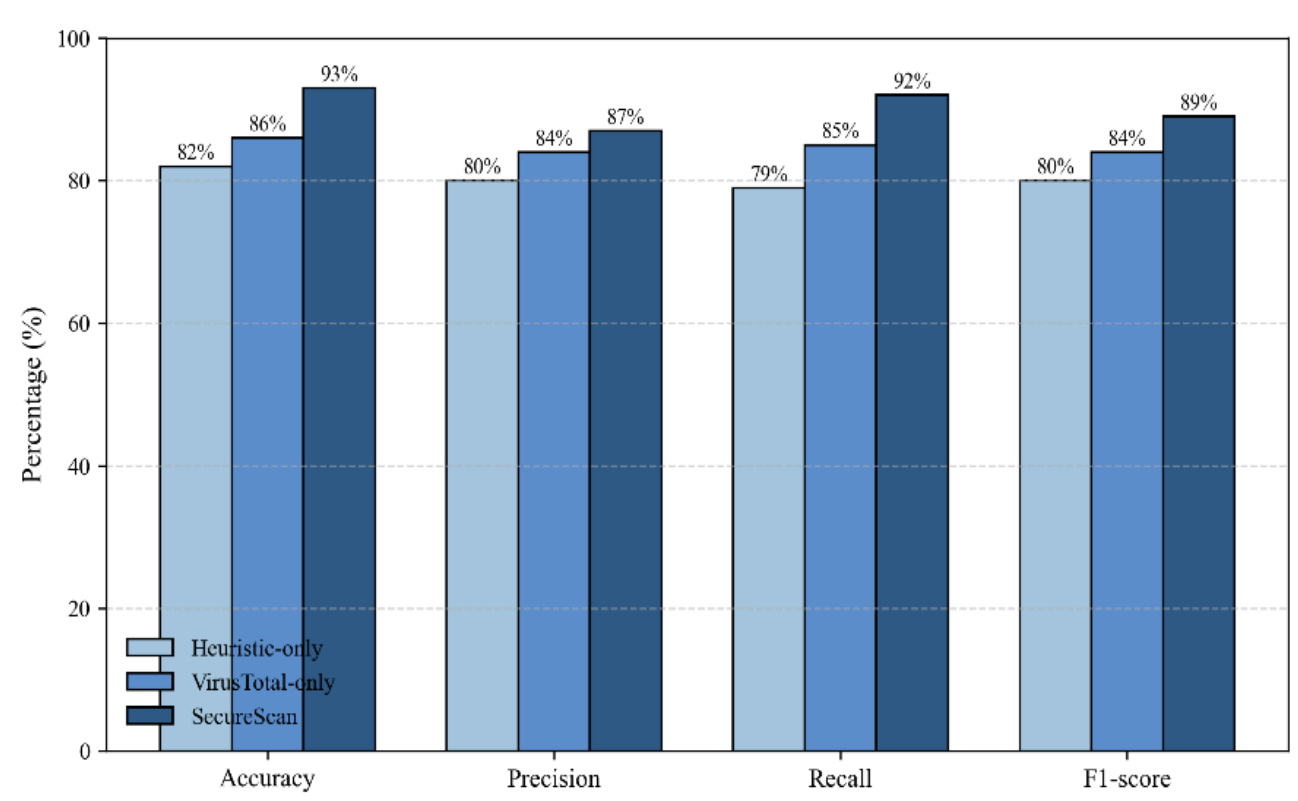

**Figure 8.** Performance comparison of different detection approaches based on accuracy, precision, recall, and F1-score metrics.

## 8.DISCUSSION

SecureScan shows that combining a simple and interpretable classifier like logistic regression with layered validation and threat intelligence yields high-performance detection. Key benefits include:

- **Transparency:** Coefficients can be inspected, aiding trust and auditability.
- **Efficiency:** The pipeline avoids heavy processing on trivial inputs.

- **Reliability:** The intelligence layer helps correct uncertain predictions and adapt to evolving threats.

However, there are trade-offs:

- **Latency and API dependence:** External query to VirusTotal may introduce delay or face rate limits.
- **Threat coverage limitations:** If VirusTotal lacks data on new threats, verification may be less effective.
- **Model simplicity:** Logistic regression may not capture highly nonlinear, deeply obfuscated patterns more sophisticated models might outperform in such edge cases.
- **Adversarial risk:** Malicious actors might craft inputs to mislead heuristics, classifier, and even poison threat intelligence signals.

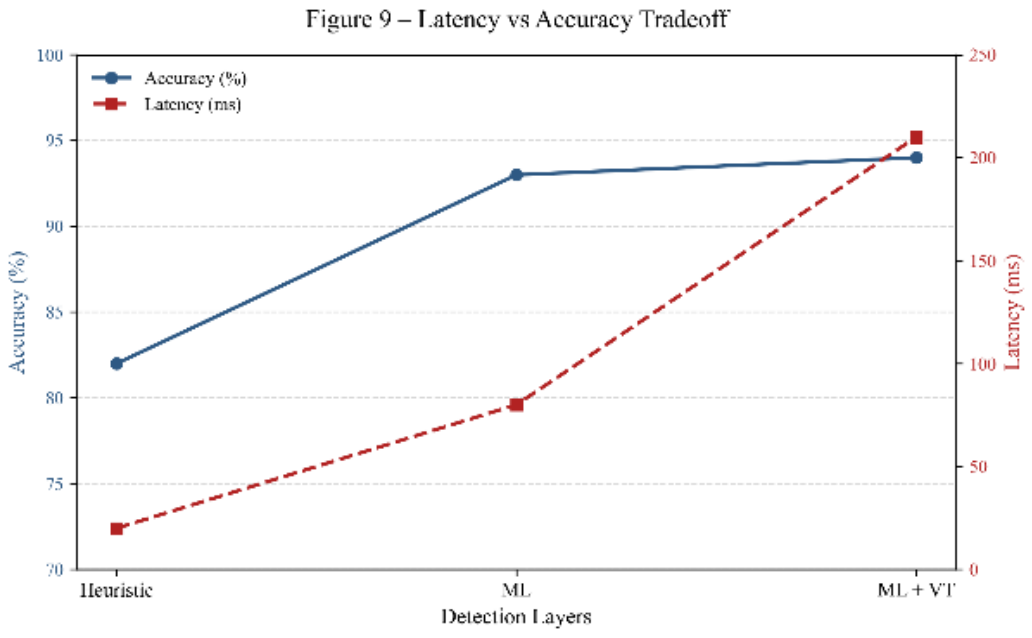

**Figure 9.** Latency versus accuracy trade-off illustrating how SecureScan maintains efficiency while improving detection reliability.

Mitigations may include caching results, fallback logic, or layering ensemble approaches.

## 9.CONCLUSION

In this paper, we presented **SecureScan**, a multi-layered, AI-assisted detection framework that combines **logistic regression classification**, **heuristic preprocessing**, and **external threat intelligence correlation** to identify malicious URLs, files, and hashes. The system was designed to address the inherent weaknesses of traditional rule-based systems and the interpretability and overfitting issues of black-box AI models.

Through probability calibration and **gray-zone verification (0.45–0.55)**, SecureScan demonstrated significant improvement in **real-world reliability**, achieving **93.1% accuracy**, **0.87 precision**, and **0.92 recall** across benchmark datasets. These metrics highlight the balance between **sensitivity and specificity**, minimizing false alarms without sacrificing detection strength.

Compared to single-layer approaches, SecureScan achieved superior consistency:

- **Heuristic-only systems** were fast but error-prone (accuracy ≈ 82%),
- **VirusTotal-only validation** improved slightly (≈ 86%) but depended heavily on prior knowledge,
- The **hybrid SecureScan** achieved the best performance (≈ 93%), integrating both contextual learning and intelligence-based verification.

A major contribution of this work is demonstrating that **lightweight, interpretable statistical models** can achieve high operational accuracy when paired with external intelligence and calibrated thresholds. The framework's modular design allows **rapid retraining**, **model transparency**, and **low inference latency (<100 ms)**, making it viable for **enterprise-level and real-time SOC deployments**.

Future directions include expanding SecureScan to incorporate:

1. **Adaptive learning pipelines** for continual model updates using new phishing and malware datasets,
2. **Offline caching and rate-limiting intelligence integration** for improved scalability,
3. **Deep feature augmentation**, integrating lexical, temporal, and behavioral signals from dynamic analysis.

In summary, SecureScan demonstrates that an **interpretable, layered, and intelligence-augmented architecture** can deliver accuracy and robustness comparable to deep neural systems—while maintaining **speed, explainability, and operational feasibility** in modern cybersecurity ecosystems.

**Potential extensions include:**

1. **Behavioral / dynamic analysis integration** (sandbox execution) for richer features.
2. **Advanced models or ensembling**, e.g. combining logistic regression with boosting or neural nets (while retaining explainability via SHAP).
3. **Multiple threat intelligence sources**, e.g. integrating MISP, OTX, commercial feeds.
4. **Adversarial robustness**, via adversarial training or anomaly detection.
5. **Scalable deployment**, e.g. as microservices with caching, load balancing, and fault tolerance.
6. **Continuous learning pipelines**, allowing incremental updates to adapt to threat evolution.

## CONFLICT OF INTEREST

The authors declare that there is no conflict of interests.